\documentclass[rmp,twocolumn,floatfix]{revtex4}

\usepackage{graphicx,amsmath,amssymb,txfonts,dcolumn}

\begin{document}

\title{Signatures of currency vertices}

\author{Petter Holme}
\affiliation{Department of Physics, Ume{\aa} University,
  901~87 Ume{\aa}, Sweden}
\affiliation{Computational Biology, Royal Institute of Technology,
  100~44 Stockholm, Sweden}

\begin{abstract}
  Many real-world networks have broad degree distributions. For some systems, this means that the functional significance of the vertices is also broadly distributed, in other cases the vertices are equally significant, but in different ways. One example of the latter case is metabolic networks, where the high-degree vertices --- the currency metabolites --- supply the molecular groups to the low-degree metabolites, and the latter are responsible for the higher-order biological function, of vital importance to the organism. In this paper, we propose a generalization of currency metabolites to currency vertices. We investigate the network structural characteristics of such systems, both in model networks and in some empirical systems. In addition to metabolic networks, we find that a network of music collaborations and a network of e-mail exchange could be described by a division of the vertices into currency vertices and others.
\end{abstract}

\maketitle

\section{Introduction}

Over the last decade methods from statistical physics have contributed greatly to the theory of complex networks~\cite{ba:rev,mejn:rev,calda:book}. One of the major contributions is the development of methods to characterize and categorize the vertices (nodes) of real-world networks. Numerous networked systems are heterogeneous in the sense that a majority of vertices have a degree lower than the average, whereas a small number of vertices have a much higher degree than the average. For many such systems, one can relate the degree of a vertex to its function. In, for example, the network of air flights~\cite{gui:air} the central vertices are the largest airports. These are the hubs international travellers hardly can avoid and arguably the most important facilities for the function of global air transportation. Degree, and other centrality measures~\cite{harary,wf}, are therefore static measures of the importance of airports to the dynamic function of the system. However, there are other networked systems with broad degree distributions where this description is incomplete. Metabolism is the set of chemical reactions occurring in a normally functioning organism. From such a reaction system, one can construct networks of chemical substances~\cite{our:curr}. Such networks have heterogeneous degree distributions. The hubs of metabolic networks are the most abundant molecules, such as CO${}_2$ and H${}_2$O. These metabolites have very different functions compared to the low-degree vertices --- they are present throughout the cell and participate in reactions of all kinds of complexity. By analogy to money, frequently changing hands, the hubs of metabolic networks are called \textit{currency metabolites}. For the overall function of the system --- to develop and maintain high-level biological functionality, and ultimately life --- low-degree vertices are also essential. Although the hubs may affect the organism's health, on average, more than the peripheral vertices, most authors agree that using degree as a proxy of functional importance is misleading~\cite{our:curr,our:bio,zhao:meta,jing:baotai,ma:meta,wagner:sw,arita:not,gui:meta}. Instead, the picture often painted is that the higher functionality, and thus the most interesting information for questions of current scientific interest (related to evolution and metabolic diseases), is contained in the organization of the non-currency metabolites. For this reason, to achieve a network that is more informative, currency metabolites are often deleted~\cite{our:curr,our:bio,zhao:meta,jing:baotai,ma:meta,wagner:sw,arita:not,gui:meta}. Another characteristic property of metabolic networks is that the non-currency metabolites form network clusters that are more connected within, than between each other. This modular structure, one believes, is related to the function of the network --- a network cluster (network module) is responsible for one relatively well-defined task in the metabolic system. The currency metabolites, on the other hand, are involved in the production of a wide variety of molecules, from many different modules. Thus the currency metabolites hide the modular network structure, something that can be used for a graph based definition of currency metabolites~\cite{our:bio}. \emph{If vertices are deleted from the network in order of highest degree, then the set of currency metabolites is the set of vertices that, if deleted, gives the highest relative modularity.} (Where ``relative modularity'' is a measure quantifying the tendency of the network to be organized in network modules, and is defined mathematically below.)

In this paper, we pursue the idea that the description of metabolic networks above --- that the bulk of the dynamics are performed by currency metabolites, and the higher order function is produced in the network modules by the low-degree vertices --- also is relevant for some other networked systems. Consider the network of people present at the venue of a larger scientific meeting, where two persons are linked with each other if they have engaged in a conversation. Probably most scientists have links to the people at the reception desk, and links to their collaborators and other scientists working on similar problems. The functional output of the conference --- the advancement of science --- would then be performed in the network clusters of people with similar interests. The receptionists, the currency vertices, are nevertheless important for the meeting to be successful, but in a different way than the other vertices. The modular structure of the scientists would be more visible if the receptionists were not included in the network. (Similar descriptions of social networks can be found in Refs.~\cite{ada:ifnw,gui:chart}.)

Whether or not a network is well described by a dichotomy of the vertices into currency and non-currency vertices is ultimately a question about the whole system, including dynamic processes on the network. Nevertheless, as mentioned above, one can define currency metabolites for any network. Since there is no general, functional definition of currency vertices one cannot evaluate the definition directly. We will perform an indirect validation by creating a model producing networks where the network characteristics of currency metabolites can be tuned continuously. Using this model, we investigate the parameter values where the designated currency vertices of the model match the identified currency vertices. By mapping out the network structure of the region in parameter space where the matching is good, one can get an indication if a network fits to the currency-vertex picture. We will also use a more direct validation for nine different types of empirical network --- we derive model parameter values from the networks and calculate the matching scores as for the model networks, a high matching will be interpreted as a support for the currency-vertex picture.

The rest of the paper is organized as follows. First, we define network modularity and currency vertices mathematically. Then, we define the network model and, finally, evaluate the currency vertices of the model and empirical networks.

\begin{figure}
  \includegraphics[width=0.8\linewidth]{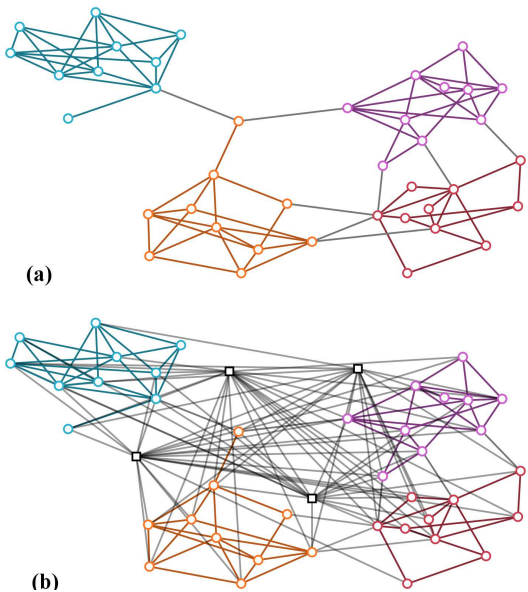}
\caption{Example output of the network model. Model parameters are $g=4$, $n_g=10$, $n_c=4$, $p_g=0.4$, $p_o=0.04$, and $p_c=0.4$. In (a) the clear modular structure of the network without the model currency vertices (MCV) is shown. In (b), we also display the MCVs obscuring the modular structure.
}
\label{fig:ill}
\end{figure}

\section{Preliminaries}

In this paper, we consider networks modelled as graphs $G=(V,E)$ where $V$ is the set of $N$ vertices and $E$ are the $M$ edges (unordered pairs of vertices). We assume the graphs to be \emph{simple}, i.e.\ that they do not have multiple edges or self-edges. (Graphs that are not simple are called multigraphs.)

\subsection{Network modularity and currency vertices}

In this section, we will discuss how to calculate network modularity. For a more detailed account, see Ref.~\cite{mejn:spectrum}. Consider a partition of the vertex set into groups, and let $e_{ij}$ denote the fraction of edges between groups $i$ and $j$. The network modularity of this partition is defined as\cite{mejn:commu}
\begin{equation}\label{eq:q}
  Q=\sum_i\left[e_{ii}-\left(\sum_je_{ij}\right)^2\right],
\end{equation}
where the sum is over all groups of vertices. The term $\left(\sum_je_{ij}\right)^2$ is the expectation value of $e_{ii}$ in a random multigraph. A prototype measure for the modularity of a graph is $Q$ maximized over all partitions, $\hat{Q}$. For many networks with broad degree distributions, it is common to measure network structure relative to a null-model of random graphs with the constraint that the set of degrees is the same as in $G$, $\mathcal{G}(G)$. In principle this means that one separates degree from other network structures, which is appropriate in our case --- in fact, this idea is implicit in the definition of currency vertices. With this null model, we subtract the average $\hat{Q}$-value for graphs in $\mathcal{G}(G)$ from $\hat{Q}(G)$:
\begin{equation}\label{eq:delta_q}
  \Delta(G) = \hat{Q}(G) - \langle \hat{Q}(G') \rangle_{G'\in
    \mathcal{G}(G)} ,
\end{equation}
where angular brackets denote average over $\mathcal{G}(G)$\cite{our:bio}. We use a random rewiring of the original graph to sample $\mathcal{G}(G)$\cite{maslov:pro}, and the heuristics proposed in Ref.~\cite{mejn:spectrum} to maximize $Q$.

To extract the currency vertices we start with the original graph $G_0$ and perform the following scheme
\begin{enumerate}
\item \label{step:q} Measure $\hat{Q}(G_i)$, where $i$ is the number of times this line has been executed before this time.
\item \label{step:del} Delete the vertex with highest degree from $G_i$ and call this graph $G_{i+1}$.
\item \label{step:cpy} Make a copy, $G_{i+1}'$, of $G_{i+1}$.
\item \label{step:swap} Rewire the edges of $G_{i+1}'$ and measure $\hat{Q}(G_{i+1}')$. Repeat this $n_{\mathrm{iter}}$ times and calculate $\langle \hat{Q}(G') \rangle_{G'\in\mathcal{G}(G)}$.
\item \label{step:condition} If $\Delta(G_i)$ is lower than $\Delta(G_0)$, or if $i=N-1$, then break the iterations.
\end{enumerate}
The vertices deleted at step~\ref{step:del} maximizing $\Delta(G_i)$ is the set of currency vertices. In this paper, we use $n_{\mathrm{iter}}=25$. A C-implementation of this algorithm can be downloaded at \url{www.csc.kth.se/~pholme/curr/}.

\subsection{Artificial networks}

To investigate the definition of currency vertices, as sketched in the Introduction, we use model networks where one can tune the strength of modularity, number of currency vertices and average degrees.

Let there be $g$ groups (corresponding to network modules), $n_g$ vertices within each group, and $n_c$ model currency vertices (MCV). Then go through all pairs of distinct non-MCVs and connect these with probability $p_g$ if they belong to the same group, and $p_o$ otherwise. Finally, go through all pairs of vertices containing at least one currency vertex and connect the pair with a probability $p_c$.

The expected number of vertices is
\begin{equation}\label{eq:n}
  N = n_c+gn_g
\end{equation}
and the expected number of edges
\begin{equation}\label{eq:m}
  M = \frac{p_ggn_g(n_g-1) + p_ogn_g^2(g-1) + p_cn_c(N-1)}{2}.
\end{equation}
The modularity $Q$ for the model with $n_c=0$ (or all MCVs removed), partitioned according to the groups, is
\begin{equation}\label{eq:q_mod}
  Q = g\frac{p_gn_g(n_g-1)}{2M}-g\left(\frac{p_gn_g(n_g-1)}{2M}+(g-1)\frac{p_on_g^2}{2M}\right)^2.
\end{equation}
In the limit $g,n_g\gg 1$, Eq.~\ref{eq:q_mod} reduces to
\begin{equation}\label{eq:q_mod_app}
Q=\frac{1}{1+g/\gamma}-\frac{1}{g}\mbox{~~where~~} \gamma=\frac{p_g}{p_o}.
\end{equation}
Since our model produces simple graphs (and not multigraphs, as the theory behind the definition of $Q$), putting $p_g=p_o$ in  Eq.~\ref{eq:q_mod}, does only approximately give $Q=0$. The error in this approximation is $\mathrm{O}(1/n_g+1/g)$. The model can easily be modified to produce multigraphs (by just dropping the requirement of no self-edges or multiple edges), in which case the $p_g=p_o$ would indeed give zero modularity.

\subsection{Matching score}

As mentioned in the Introduction, we will investigate how well the original structure of the network matches the output of the currency-vertex detection algorithm as a function of model parameter values. The quantity for measuring the overlap of model groups and identified network clusters is the fraction of overlapping group identities in the best matching between the two classifications. In other words, let $x_i$ be vertex $i$'s group in the original network ($x_i\in [1,\cdots,g]$, currency vertices are not counted as members of any group) and let $y_i$ be vertex $i$'s identity obtained from the currency-vertex detection ($y_i\in [1,\cdots,N_g]$, $N_g$ is the number of detected groups). Then find the labeling of the graph-clustering groups such that each group has a unique number in the interval $[1,\cdots,N_g]$, and that the number $n_{\mathrm{match}}$ of vertices $i$ with $x_i=y_i$ is maximized. Then we define the \textit{matching score} $\mu_g = n_{\mathrm{match}} / gn_g$. We calculate $n_{\mathrm{match}}$ by a simple heuristic:
\begin{enumerate}
\item \label{step:start} Start with a random labeling of the groups.
\item \label{step:chose} Select a pair of group labels.
\item \label{step:shift} If $n_{\mathrm{match}}$ does not decrease if these labels are swapped, then swap them.
\item \label{step:rep1} If no improvement has been made during the last $n_{\mathrm{rep}}$ steps, go to step~\ref{step:chose}.
\item \label{step:rep2} Start over from step~\ref{step:start} with a new random seed unless a new highest $n_{\mathrm{match}}$ has been found in step~\ref{step:rep1} the last $N_{\mathrm{rep}}$ time steps.
\end{enumerate}

In addition to measuring the matching of model groups and network clusters, we look at the matching between actual currency vertices (identified by the algorithm), and the MCVs assigned in the model during the generation of the graph. In this case, we use the Jaccard index of the two sets of vertices:
\begin{equation}\label{eq:jaccard}
  \mu_c = \frac{|V_c\cap V_C|}{|V_c\cup V_C|},
\end{equation}
where $V_c$ is the set of detected currency vertices, $V_C$ is the set of MCVs, and $|\;\cdot\;|$ denote the number of elements of a set.

\section{Numerical results}

\begin{figure}
  \includegraphics[width=0.8\linewidth]{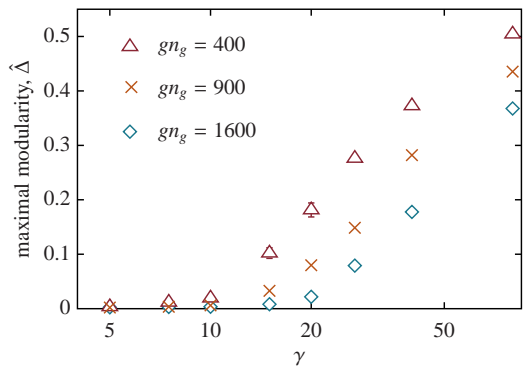}
\caption{ The maximal relative modularity as a function of the ratio $\gamma=p_g/p_o$ of probabilities for attachment within a group.  We chose $n_c=10$, $g=n_g$ and other parameter values such that the average degree is $181.1$ for model currency vertices, and $14.5$ for the others. The points are averages of $10$ to $20$ network realizations.
}
\label{fig:mod1}
\end{figure}

\begin{figure}
  \includegraphics[width=0.8\linewidth]{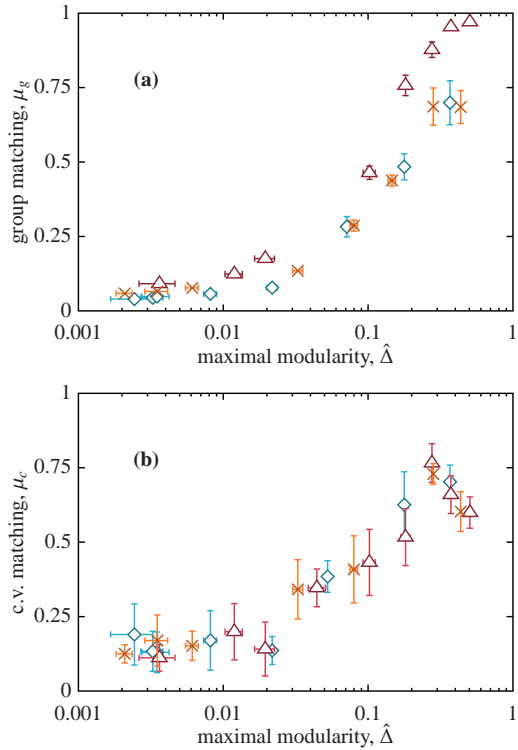}
\caption{Matching scores for networks of different sizes. (a) shows the group matching scores $\mu_g$. (b) displays the currency-vertex matching scores. The symbols and parameter values are the same as in Fig.~\protect\ref{fig:mod1}. 
}
\label{fig:mod2}
\end{figure}

\begin{figure}
  \includegraphics[width=0.8\linewidth]{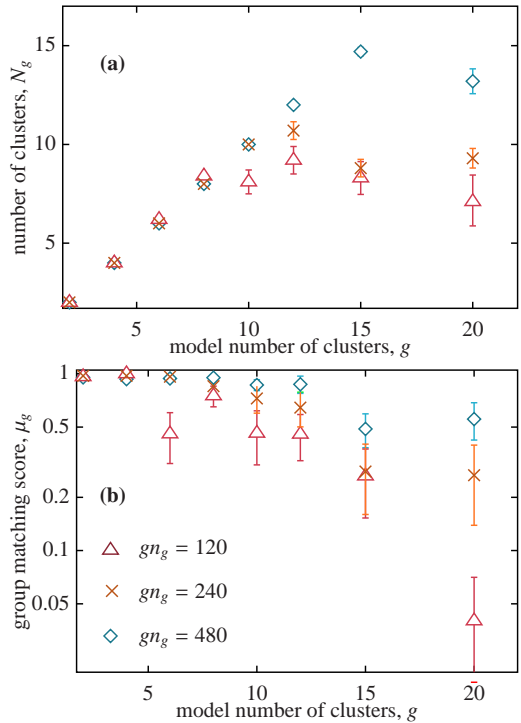}
\caption{The number of network clusters $N_g$ as a function of the number of groups $g$ in the model (a), and the group matching score $\mu_g$ as a function of $g$ (b). The other parameter values are $p_g=0.2$, $p_o=0.01$, $p_c=0.25$ and $n_c=10$. Averages are over $20$ network realizations. 
}
\label{fig:ng}
\end{figure}

\begin{figure}
  \includegraphics[width=0.8\linewidth]{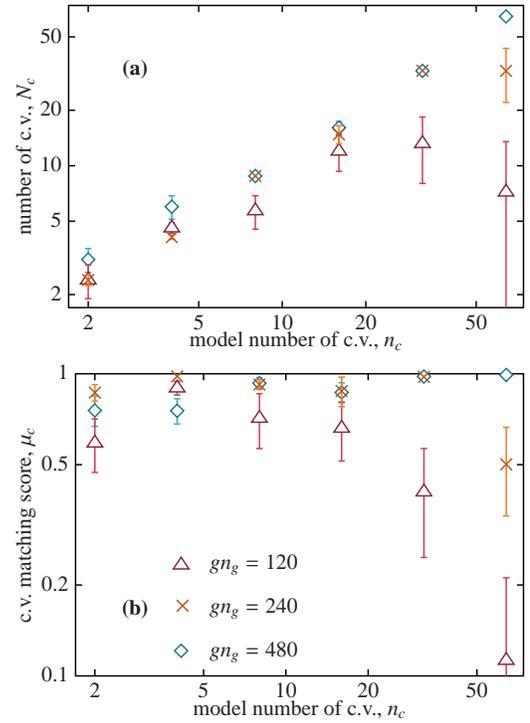}
\caption{ The number of currency vertices $N_c$ as a function of the number of model currency vertices $n_c$ (a), and the currency-vertex matching score $\mu_c$ as a function of $n_c$ (b). The other parameter values are $p_g=0.2$, $p_o=0.01$, $p_c=0.25$ and $g=6$. Averages are over $20$ network realizations. 
}
\label{fig:nc}
\end{figure}

\subsection{Artificial networks}

We start our numerical investigation by measuring the matching scores for networks of different modularity. As hinted from Eq.~\ref{eq:q_mod}, the modularity can be controlled by the ratio of edges between vertices of the same, and different, groups $\gamma$. The measurable modularity (i.e.\ the one that does not need the partition information from the network construction) is (for fixed network sizes) monotonously increasing with $\gamma$, see Fig.~\ref{fig:mod1}. This confirms the indication from Eq.~\ref{eq:q_mod_app} that $\gamma$ works as a control parameter for the relative modularity. We also see that the maximal value of the relative modularity $\hat\Delta$ depends on both the network size and $\gamma$. This effect is smaller if one let the degree increase with the number of vertices (which as been observed in some classes of networks\cite{doro:acc,pok}), instead of keeping degree fixed as in Fig.~\ref{fig:mod1}. This also suggests that comparing the $\Delta$-values of different networks should be done carefully. The comparison built into the currency-vertex definition algorithm concerns a sequence of monotonously shrinking networks from the same original. Since the size of the network do not change much during an iteration, and due to the smooth monotonous increase of Fig.~\ref{fig:mod1}, the shrinking size during the currency-vertex definition scheme is not a technical problem.

For real-world networks the $\hat\Delta$ (and not $\gamma$) is a measurable quantity. In Fig.~\ref{fig:mod2}, we show the matching scores as function of maximal relative modularity $\hat\Delta$. The matching scores (both $\mu_g$ and $\mu_c$) increase monotonously with $\hat\Delta$, meaning that the picture of regular vertices grouped into clusters (instantiated by the model) holds better the larger the relative modularity is. For the parameter values in question, $\hat\Delta$-values of $\sim 0.2$ are needed for matching-score values over $0.5$. For example, if one deems values of $\mu_g$ and $\mu_c$ less than $0.5$ too small, then one can conclude that networks with $\hat\Delta<0.2$ probably do not fit the currency-vertex description. We note that in Fig.~\ref{fig:mod2}, the matching scores for a given $\hat\Delta$-value seem to converge from above. If the $p$-parameters ($p_g$, $p_o$ and $p_c$) are fixed as $N$ are changed, then this convergence goes in the opposite direction (the $\mu$-values grow with the system size).

In Fig.~\ref{fig:ng}, we investigate how the number of network clusters $N_g$ depends on the number of groups $g$ in the model. For a small number of groups $g\approx N_g$ (as seen in Fig.~\ref{fig:ng}(a)). Indeed, the identified clusters are almost the same as the original groups ($\mu_g\approx 1$ in Fig.~\ref{fig:ng}(b)). For larger $g$, $N_g$ starts to deviate from $g$. This deviation appears later for larger network sizes, indicating that this is a finite size effect. The number of vertices sets a (trivial) upper bound of this matching. Fig.~\ref{fig:ng}(a) shows that the bound increases slower than linear (possibly logarithmically). In the light of this observation, if $N_g$ is too large (considering the network sizes), then the currency-vertex picture seems less appropriate.

Fig.~\ref{fig:nc} illustrates the model's dependence of the number of MCVs, $n_c$. Just like for the number of network clusters, the matching with the corresponding model parameters is largest for small values. For larger values of $n_c$, the number of currency vertices start to deviate (becoming lower than $n_c$). From both the $N_c$- and $\mu_c$-curves, we note that matching score is larger for larger networks. The mismatch between the currency vertices of the model network construction and the currency vertices by definition is thus a finite-size effect.

\begin{table*}\label{tab:empir}
\begin{ruledtabular}
\begin{tabular}{r|c|rrrrrrrrr}
network & Ref. & $N$ & $M$ & $N_c$ & $N_g$ & $\hat{n}_g/N$ & $\Delta_0$ & $\hat\Delta$ & $\mu_g$ & $\mu_c$ \\\hline
music collaborations & \cite{jazz} & 198 & 2256 & 16 & 5 & 0.273 & 0.261 & 0.318 & 0.98(1) & 0.81(7) \\
metabolic & \cite{our:curr} & 473 & 1694 & 4 & 13 & 0.214 & 0.303 & 0.349 & 0.80(2) & 0.68(9) \\
e-mail & \cite{bornholdt:email} & 1133 & 5161 & 10 & 15 & 0.286 & 0.189 & 0.247 & 0.44(6) & 0.45(9) \\
protein interaction & \cite{hh:pfp} & 4168 & 7434 & 13 & 41 & 0.108 & 0.080 & 0.099 & 0.05(5) & 0.07(4) \\
airport network & \cite{my:cps} & 456 & 2799 & 29 & 24 & 0.283 & 0.128 & 0.184 & 0.05(3) & 0.065(2) \\
neural network & \cite{cenn:brenner} & 280 & 1973 & 32 & 6 & 0.257 & 0.186 & 0.232 & 0.29(4) & 0.02(1) \\
dolphin social network & \cite{dolph} & 62 & 159 & 0 & 4 & 0.339 & 0.166 & 0.166 & 0.85(2) & -- \\
atmospheric & \cite{sole:astro} & 249 & 1197 & 0 & 4 & 0.518 & 0.122 & 0.122 & 0.33(1) & -- \\
software dependence & \cite{mejn:mix} & 1033 & 1718 & 0 & 29 & 0.181 & 0.148 & 0.148 & 0.19(1) & -- \\
\end{tabular}
\end{ruledtabular}
\caption{Values (network sizes, number of currency vertices $N_c$, number of network clusters $N_g$, relative size of the largest cluster $\hat{n}_g/N$, relative modularity $\Delta_0$ of the original network, maximal relative modularity $\hat\Delta$, group matching score $\mu_g$, currency-vertex matching score $\mu_c$) for empirical networks. In the music collaboration network, vertices are jazz musicians, connected if they have appeared on the same recording. In the metabolic and atmospheric networks the vertices are chemical substances and edges represent pairs of substances participating in the same reaction. The metabolic data comes from reactions in the bacterium \textit{Mycoplasma genitalium} and the atmospheric data regards Earth. In the e-mail network, vertices are e-mail addresses and edges mean that at least one e-mail within the three month sampling period has been sent from one address to the other. The protein interaction network consists of proteins connected if they can bind physically to one another. Vertices in the neural network are neuronal cells of the nematode \textit{Caenorhabditis elegans}, and edges indicate how these are connected. In the airport network, vertices are North American airports and edges pairs of airports with a regular nonstop flight. The dolphin social network is based on observed interactions between bottlenose dolphins in Doubtful Sound, New Zealand. In the software dependence data, a vertex is a software package and a link indicate that one package requires another package to be installed to function. Some of these network datasets are originally directed (the neuronal and e-mail networks are also weighted). These are transformed into simple graphs by reciprocating directed edges and treating any non-zero, weighed edge as an unweighted edge. The table is ordered primarily according to the $\mu_c$-values, secondarily after the $\mu_g$-values. The numbers in parentheses are standard errors in units of the last decimal.}
\end{table*}

\subsection{Empirical networks}

Now we turn to evaluate real-world networks. We perform the identification of currency vertices as outlined above, and obtain a decomposition into network clusters of the non-currency vertices. From this we obtain values of $N_c$, $N_g$, $\Delta_0$ and $\hat\Delta$ displayed in Table~\ref{tab:empir}. Furthermore, we calculate $\mu_c$ and $\mu_g$ for our model with parameter values derived from the network --- we let $g$ be the measured $N_g$, set $n_c$ equal to $N_c$, $n_g=(N-N_c)/g$ (rounded to the lower integer) and, for $p_g$, $p_o$ and $p_c$, use the fraction of edges between the respective types of vertices in the empirical network. By this procedure, we obtain matching scores giving some indication how appropriate the currency-vertex picture is. One difference between the model and the empirical network is that the clusters of the model have the same sizes, whereas the cluster sizes of the real-world network varies. This is a feature that could affect the results quantitatively, especially if there is a wide distribution of cluster sizes. This is (fortunately for the analysis method) not the case. Even if the degree distributions are broad, the cluster size distribution is rather narrow --- the $\hat{n}_g/N$ values of Table~\ref{tab:empir} are low, with the atmospheric network as an exception (the results for this network thus be taken with a grain of salt).

Of the nine empirical networks, three networks do not have any currency vertices at all. These three are clearly disqualified for our currency-vertex picture. Of the six networks with $N_c>0$, three networks --- a social network of music collaborations, a metabolic network and a network of e-mails --- have larger $\mu_c$- and $\mu_g$-values than other networks. These networks fulfil the structural prerequisites for a currency-vertex picture. In the music collaboration network, we can assume the currency vertices are studio musicians that are not strongly affiliated with one group, or orchestra, but participate on many artists' recordings. The e-mail network does not include spam mails\cite{bornholdt:email}, so we assume the hubs are addresses that send, or receive, information of more general nature (cf.\ the example of the social interactions at a scientific meeting in the Introduction). We also note that this classification seems independent of the network sizes --- of the three networks with large matching scores (and $n_c>0$), the collaboration network is comparatively small and dense, whereas the e-mail network is larger and sparser; also among the networks with low $\mu$-values, this observation holds (the protein interaction network is large and sparse, the neural network is denser and smaller). Furthermore, we note that the region of the network-structure space (for $\hat\Delta$, $N_g$ and $N_c$) giving large matching scores (as found in the previous section) is consistent with the observations in Table~\ref{tab:empir}. Examples of networks falling outside of these ranges are the airport network (with a too large $N_g$-value considering its size), and the neural network (having too many currency vertices for its size to have a good matching).

\section{Conclusions}

In this paper, we have extended a organizational principle, known in metabolic networks, to networks in general. In this picture, most vertices are of relatively low-degree, grouped into relatively distinct network clusters. A small minority of the vertices, however, have much larger degree than the average, are linked to vertices of all clusters, and thereby obscure the modular organization of the low-degree vertices. We call these currency vertices. In a functional interpretation of this picture, the currency vertices perform the bulk of the dynamics, whereas the more specialized (and not necessarily less important) features of the system occur in the modules.

By just measuring the modular structure of a network, one cannot validate the currency-vertex definition. Instead of a direct validation, one can assume the network itself is an encoding of the functions of the vertices, and the currency-vertex definition is a decoding of this information\cite{rosvall:maps,mejn:spectrum,mejn:mix}. Following this philosophy, we create a model with a tunable number of currency vertices, number of network clusters and strength of these features. The match between the encoded and decoded sets of currency vertices and network clusters are closest if the modularity is large, and numbers of currency vertices and network clusters are low. Using this procedure, we also evaluate empirical networks. We conclude that three of nine investigated networks fit rather well to the currency-vertex picture. The first of these networks is a network of collaborations between music artists, where we assume the currency vertices are studio musicians and the other vertices are group, or band, members (and the network clusters are the music groups). Our second example of a network with currency-vertex structure is a metabolic network --- appropriate, since this class of networks is the inspiration of the concept. The third network potentially fitting our picture is an e-mail network, where we interpret the currency vertices as senders, or receivers, of general content e-mails (since the e-mails are sampled from a group of university e-mail accounts, such e-mails could be information to and from the university administration). The dialogues between colleagues and classmates presumably take place within the network clusters. These dialogs correspond to a different type of information process than the e-mails to the hubs, just as the function of currency metabolites is different from other substances in metabolic networks and the hubs of the music collaboration network have different roles than the majority of musicians. Among the networks not fitting the picture of currency vertices are a social network of dolphins (with a clear modular structure, but no currency metabolites), a network of airports and a network derived from chemical reactions in the Earth's atmosphere.

We have described the currency metabolite picture as a dichotomous property --- networks either fit it, or not. This is just a simplification and one may argue that the hubs of e.g.\ the airport networks (if we for a moment ignore that our airport network did not pass our tests) share some of the characteristics of currency vertices in other networks. At least, larger airports have a larger fraction of transfer passengers, and thus a somewhat different function in the entire dynamic system of air travel. This also illustrates that, to determine how well characterized a network is by a division of the vertices into currency vertices and others, one needs to (in addition to the analysis presented in this paper) consider the dynamics of the subject system.

\section*{Acknowledgment}

P.H. acknowledges economic support from the Swedish Foundation for Strategic Research and thanks Holger Ebel, Michael Gastner, Mikael Huss, Andreea Munteanu and Mark Newman for data.

\end{document}